# On the Role of Quantum Computing in Science and Cybersecurity

Giovanny Espitia and Jungkyu Park, P.h.D

Kennesaw State University

## ABSTRACT

In this paper, we examine the state art of quantum computing and analyze its potential effects in scientific computing and cybersecurity. Additionally, a non-technical description of the mechanics of the listed form of computing is provided to educate the reader for better understanding of the arguments provided. The purpose of this study is not only to increase awareness in this nescient technology, but also serve as a general reference guide for any individual wishing to study other applications of quantum computing in areas that include finance, bio – chemistry, and data science. Lastly, an educated argument is provided in the discussion section that addresses the implications this form of computing will have in the main areas examined.

## I. INTRODUCTION

In latter half of the 1930's, two prominent mathematicians hypothesized the idea of a universal machine capable of simulating any other machine [3]. This type of machine came to be known as "Universal Turing Machine" in honor of its main postulator, *Alan Turing*. Such machine was soon realized, giving rise to the information age we live in. Additionally, the realization of such machine allowed for rapid progress in science and geopolitics along with related fields to the former disciplines. For this reason, our civilization has benefited from unprecedented wealth and a high standard of living relative to our species history. However, based on the laws of physics, the exponential increase that has been commonplace up this stage is bound to halt as advances in microchip design approach the single atom scale. Interestingly, a solution to this problem was proposed in the latter half of the 1980's by physicist and Nobel laurate, *Richard Feynman* [4]. This solution calls for an entirely new computational paradigm known as quantum computing. In this paper, a general description of this form of computing will be provided in an attempt to answer the question, how will quantum computing affect our perception of modern science and geopolitics in terms of computational models and cybersecurity?

## II. TECHNOLOGY OVERVIEW

Quantum computing is a computational paradigm that makes use of different phenomena developed in the theory of quantum mechanics. These phenomena allow for the existence of a different unit of information known as the quantum bit or "qubit" for short. In contrast to the classical bit, a qubit makes use of three fundamental principles that include superposition, interference, and entanglement. These three principles endow the discussed qubit with the property of spin capable of existing in a superposition of states as opposed to a binary on/off state experienced by classical systems. In other words, if a classical system can perform $n$ operations, a quantum system can perform $2^n$ logic operations [1]. This capability translates into an array of applications, including scientific computing and cybersecurity.

## III. APPLICATIONS AND CAPABILITIES

Quantum computing opens a wealth of possibilities for scientists and engineers to better examine and study their models. Over the last couple of decades, scientist begin to employ computational programs to better understand real world phenomena as opposed to the traditional mathematical models. One example being "The Wolfram Physics Project" [11] led by *Wolfram et al*. Furthermore, this computational approach has allowed for safer and more efficient engineering by allowing numerical approximations to systems of differential equations, whose solutions cannot be obtained by analytical methods. The latter accomplishment has allowed for computer aided design (CAD) software to decrease the amount of time required to design a system, while at the same time increasing its effectiveness and safety grade. However, as scientists and engineers begin to study more complex phenomena in a linear fashion, sooner rather than later, our silicon-based computing infrastructure will not be able to meet these rigorous demands. For this reason, quantum computers no longer become a luxury, but rather a necessity. Quantum computers as mentioned before, are capable of completing a task of comparable complexity to that of a classical computer not only with a lower number of digits of information, but also with a smaller wastage of natural resources such as electricity. Furthermore, quantum computers have less difficulty in simulating interactions at the atomic level relative to a classical system, in part, because the system running the simulation is a quantum system by itself.

This translates into even more sophisticated applications including molecular dynamics, protein folding, and enzyme simulation for the development of new medicine. For this reason, various academic, private, and governmental institutions have begun research efforts in this discipline at an unprecedented level of support in our history.

Besides from these scientific applications, quantum computing is recognized by its revolutionary potential in cryptography and thus cybersecurity. As our society has transitioned to an economy filled and essentially run by data, information, and computing systems, the proper protection of the latter becomes essential. Up to this stage, modern cybersecurity systems are based in a paradigm known as RSA [8] encryption. This form of encryption attributes its success to the computational complexity associated with the factorization of large prime numbers. The latter represents a significant challenge to modern classical computers, but a quantum computer programmed with an algorithm known as Shor's algorithm [10] is able to achieve this computation in a fraction of time.

## IV. REALIZATION EFFORTS

The race toward seamless quantum computers begun in the latter part of the 2010's with private companies including IBM, Google, and D- Wave as the dominant figures. The efforts conducted by the latter has already allowed computations that cannot be performed in a classical computer in a reasonable amount of time. One example is Google's Sycamore processor being able to perform a calculation in 200 second as opposed to 10, 000 years in a classical system [7]. In recent years however, this western domination has begun to shift toward south eastern countries, particularly China. The latter can be attributed to significant governmental support with the primary goal to assert leadership in this field.

## V. DIFFERING APPROACHES

The achievement of the studied technology is based on three criteria. These include the funding allotted toward this effort, the political ideologies, and the state of the art of established channels of communication.

In the United States, complying with the established ideology of capitalism, the main entities striving for this technology are private companies as opposed to national laboratories or universities. This fact is analogous with NASA outsourcing the design and manufacturing of the various spacecraft systems during the Gemini and Apollo programs as opposed to conducting the listed processes in house. In this race however, the main companies performing the effort for the United States include Google, IBM, and ION Q. These companies have taken a different approach toward this technology that is optimized for a particular task. For example, Google's subdivision AI Quantum has opted for an architecture known as "Quantum Turing Machine" [5] that could theoretically serve as a universal computing machine. This type of architecture will enable a quantum system to function in a similar way to that of a classical system. In contrast to this, IBM is striving toward an architecture known as "Adiabatic" [9] capable of outperforming the architecture pursued by Google in the short term. This occurs, because Adiabatic systems are configured for a particular task and are intended to solve optimization problems, thus finding most of their users in academia. Lastly, ION Q which is a startup company founded by two former professors from Maryland and Duke, is striving for a trapped ion architecture [6]. This architecture was first proposed by the academics referred and it has the potential to outperform physical qubits across different computations including simulation, time to coherence, and predictability.

In contrast to the Unites States, The People's Republic of China has made the realization of this technology into a task of national priority. The latter has been indicated in three consecutive 5 – year plans devised by the government. Additionally, China's national secretary, Xi Jinping, has listed the advancement of quantum information science as a must due to it having the potential to cause major repercussions in both the national security and cybersecurity spaces. To achieve this technology, China is relying primarily in government funded institutions including the Chinese Academy of Sciences, the National Science Foundation of China, and other research institutions including Tsinghua and Pekin Universities. Furthermore, differing from the competitive environment created by capitalism in the United States, China plans to advance this technology in a united and cooperative fashion that has already begin to prove its worth. For example, to combat cybersecurity attacks, a 4, 600 km quantum network [2] has already been built to provide an alternative path toward the flow of quantum keys and thus a theoretically uncrackable network of communication. In addition, to complement the network described, in 2016 China launched the first quantum satellite known as "Micious" capable to communicating with ground bases following the Quantum Key distribution paradigm [12]. The latter network has been devised with the primary intent to enhance established channels of communication, but it has the potential to achieve mass adoption through what is called "The Quantum Internet". Theoretically, this type of internet is superior to the established one due to its use of Quantum Key Distribution, thus making it practically uncrackable.

**VI. DISCUSSION**

Now that a better understanding of quantum computing as a scientific and national security tool is present, one can venture to the main premise of this paper. So how will quantum computing affect our understanding of computational science?

Quantum computing will increase the capabilities researchers have when it comes to simulations with the objective of modeling their hypothesis or query. Due to the nature of the systems previously discussed, the reliability of results gathered will be higher and thus indistinguishable from the real system. This fact will force current nay-sayers against quantum computing to re-consider the opportunities presented by this technology in terms of study. For example, a cosmologist could simulate the event horizon that occurs in a black hole for any arbitrary number of black holes. With this example, one can appreciate the increase in efficiency in terms of time, due to the cosmologist being able to simulate simultaneous models as opposed to one in the same amount of time. As a result, computational science will shift from being the last resort for experimentation to the primary one.

Next, how will quantum computing affect the way geopolitics is conducted? Towards the end of the 20th century, with the drastic progress in computer technology, cybersecurity and protection of encrypted information became more prominent than ever. In addition, the introduction of public key cryptography and RSA encryption, led many experts in the area to conclude that such field was completed. However, as mentioned before, mathematician Peter Shor [10] developed an algorithm capable of performing prime factorization and thus posing a threat not only to the RSA paradigm, but any organization that made usage of this theory to encrypt their information. For this reason, an unprecedented interested from newly dominant figures in the world stage became evident. China began a national effort to realize this technology in a speedier manner than their American counterparts. With great investments in terms of physical and human capital, China is close to their goal and thus pose a threat to their political rivals including the United States. This event will shift the manner in which geopolitics is conducted by inducing a transition in the resource of choice from oil and nuclear reserves, to information and top governmental secrets.

**CONCLUSION**

Quantum computing is a revolutionary paradigm that combines the main ideas developed in 20th century. From quantum mechanics to information theory, from computational cryptography to cybersecurity. This technology has been devised to stay, but its effects in society are yet to be seen. Nonetheless, with a working knowledge and competency in history, one can provide educated arguments for what the future holds as provisioned in this paper.

**REFERENCES**

1] Bernhardt, C. (2020). *Quantum Computing for Everyone (Mit Press)* (Illustrated ed.). The MIT Press.

[2] Chen, Y. A., Zhang, Q., Chen, T. Y., Cai, W. Q., Liao, S. K., Zhang, J., Chen, K., Yin, J., Ren, J. G., Chen, Z., Han, S. L., Yu, Q., Liang, K., Zhou, F., Yuan, X., Zhao, M. S., Wang, T. Y., Jiang, X., Zhang, L., . . . Pan, J. W. (2021). An integrated space-to-ground quantum communication network over 4,600 kilometres. *Nature*, *589*(7841), 214–219. https://doi.org/10.1038/s41586-020-03093-8

[3] Copeland, J. (2007). The Church-Turing Thesis. *NeuroQuantology*, *2*(2). https://doi.org/10.14704/nq.2004.2.2.40

[4] Feynman, R. P. (1982). Simulating physics with computers. *International Journal of Theoretical Physics*, *21*(6–7), 467–488. https://doi.org/10.1007/bf02650179

[5] Muller, M. (2008). Strongly Universal Quantum Turing Machines and Invariance of Kolmogorov Complexity. *IEEE Transactions on Information Theory*, *54*(2), 763–780. https://doi.org/10.1109/tit.2007.913263

[6] Pino, J. M., Dreiling, J. M., Figgatt, C., Gaebler, J. P., Moses, S. A., Allman, M. S., Baldwin, C. H., Foss-Feig, M., Hayes, D., Mayer, K., Ryan-Anderson, C., & Neyenhuis, B. (2021). Demonstration of the trapped-ion quantum CCD computer architecture. *Nature*, *592*(7853), 209–213. https://doi.org/10.1038/s41586-021-03318-4

[7] *Quantum Supremacy Using a Programmable Superconducting Processor*. (2019, October 23). Google AI Blog. https://ai.googleblog.com/2019/10/quantum-supremacy-using-programmable.html

[8] Rivest, R. L., Shamir, A., & Adleman, L. (1983). A method for obtaining digital signatures and public-key cryptosystems. *Communications of the ACM*, *26*(1), 96–99. https://doi.org/10.1145/357980.358017

[9] Saravanan. (2014). NOVEL REVERSIBLE VARIABLE PRECISION MULTIPLIER USING REVERSIBLE LOGIC GATES. *Journal of Computer Science*, *10*(7), 1135–1138. https://doi.org/10.3844/jcssp.2014.1135.1138

[10] Shor, P. W. (1999). Polynomial-Time Algorithms for Prime Factorization and Discrete Logarithms on a Quantum Computer. *SIAM Review*, *41*(2), 303–332. https://doi.org/10.1137/s0036144598347011

[11] *Wolfram Physics Project | A Class of Models with the Potential to Represent Fundamental Physics*. (2020). Wolfram Physics Project. https://www.wolframphysics.org/technical-introduction/

[12] Yuen, H. P. (2016). Security of Quantum Key Distribution. *IEEE Access*, *4*, 724–749. https://doi.org/10.1109/access.2016.2528227